\documentclass[10pt, conference, letterpaper]{IEEEtran}
\usepackage{booktabs} 
\usepackage{color,soul}
\usepackage[T1]{fontenc}
\usepackage[utf8]{inputenc}
\usepackage{balance}
\usepackage[printonlyused]{acronym}
\usepackage{xcolor}
\usepackage{comment}
\usepackage{amsmath}
\usepackage{subfig}
\usepackage{gnuplottex}

\usepackage[font=small, labelfont=bf, belowskip=-1pt, aboveskip=1pt]{caption}

\ifCLASSOPTIONcompsoc \else \fi \ifCLASSINFOpdf \else \fi
\begin{document}
\title{{Internet-Scale Video Streaming over NDN}}
\author{\IEEEauthorblockN{Chavoosh Ghasemi$^1$, Hamed Yousefi$^2$, Beichuan Zhang$^1$}
	\IEEEauthorblockA{$^1$University of Arizona, USA\\$^2$Aryaka Networks, USA
	}
}
\maketitle
\begin{abstract}

Research in Information-Centric Networking (ICN) and Named Data Networking (NDN) has produced many protocol designs and software prototypes, but they need to be validated and evaluated by real usage on the Internet, which is also critical to the realization of the ICN/NDN vision in the long-run. This paper presents the first Internet-scale adaptive video streaming service over NDN, which allows regular users to watch videos using NDN technology with no software installation or manual configuration. Since mid-2019, the official NDN website~\cite{ndn-website} has started using our service to deliver its videos to Internet users over the global NDN testbed, showcasing the feasibility of NDN. We conduct real-world experiments on the NDN testbed to validate the proper implementation of the client, network, and server sides of the proposed system and also evaluate the performance of our video streaming service.


\end{abstract}
\vspace{0.5cm}
\begin{IEEEkeywords}
Adaptive Bit-Rate Streaming, Named Data Networking, Video Streaming over HTTP.
\end{IEEEkeywords}

\section{Introduction}
Significant changes in Internet usage have led to the rapid development of information-centric networking (ICN) architectures, such as Named Data Networking (NDN)~\cite{Zhang14}\cite{Ghasemi'19}, which makes a fundamental shift from address-centric to content-centric communications. Over almost a decade of research, many protocols, mechanisms, and software prototypes have been designed and implemented. Still, they have often been evaluated in simulations or small-scale demonstrations. At this stage of architectural development, what is keenly needed is a trial deployment on the Internet that can attract real users and generate real traffic. Such a deployment will demonstrate NDN's feasibility in real-world settings and its benefits to applications while providing necessary data to validate and evaluate existing designs and implementations. 
We direct the main focus of this paper toward realizing an adaptive video streaming service entirely over NDN while deploying it over the global NDN testbed for general Internet users. We choose video streaming as it is one of the most popular applications on the current Internet, contributing to the majority of modern Internet traffic, and thus offering a better chance to attract users and traffic. (According to the Cisco VNI report~\cite{CISCO_VNI}, video traffic will account for 82\% of total Internet traffic by 2022.) 

Challenges in deploying NDN services in the wild are two-folds: usability and performance. General NDN deployment requires software installation and configuration at end-hosts (Sec. II), which many users do not want to or are not capable of performing. Besides, most existing NDN applications are developed from scratch with entirely different user interfaces, which can make users not comfortable to use them. When users actually use the service, they would expect Quality of Experience (QoE) comparable to that of popular services they have accustomed to; 
otherwise, they may give up on using the NDN service. Therefore, to successfully attract general Internet users and real traffic, the deployed service needs to be transparent to end-users and have reasonably good performance.



In this paper, we develop an adaptive bit-rate video streaming service over NDN by employing the NDN community’s rich collection of libraries and tools, developed and validated over almost a decade, and open-source community (Sec. III). Since mid-2019, the NDN website has started to use our video streaming service instead of third-party services, like Youtube and Vimeo, to deliver its video contents to regular Internet users over NDN protocol. This also allows us to validate and evaluate our system over the NDN testbed by streaming videos of NDN website (Sec. IV). The results confirm that: (1) video delivery adapts to the highest usable quality based on network conditions, (2) the overall performance is satisfactory as measured by delays, jitter, and user QoE, (3) in-network caching improves the streaming performance. This work also shows the feasibility of NDN technology to operate in the wild with real-world settings. The experiments, however, reveal a number of interesting open problems on in-network caching to investigate in the NDN community (Sec. V). We open-sourced our software and published its demos and tutorials on the project's website~\cite{ivisa-website} to incentivize developing more such services/applications over NDN.





\section{Background and Related Work}

\subsection{Basics of Video Streaming}
After HTML5 standardization, video streaming over HTTP has rapidly become popular as browsers could playback an embedded video with no need for any plugin. 
To improve QoE and maximize connection utilization, ``adaptive bit-rate'' video streaming has been proposed. Adaptiveness is vital to today's streaming services for supporting various types of end-user devices with different hardware/software resources. To support adaptive bit-rate feature, (1) the server needs to prepare a set of different versions (called \textit{representations}) of each video file, each of which with different characteristics (e.g., frame size, bit-rate, codec, etc.) and package them according to a packaging standard such as HTTP Live Streaming (HLS). The packaging process divides each representation into a series of small segments, each of which contains a few seconds of video, and generates the playlist files to reference the segments, and 
(2) the browser on the end-user side needs to decide which representation to retrieve from the server for a smooth but high quality streaming by actively monitoring its network and processing resources. The modern browsers, however, do not natively support bit-rate adaptation logic, so third-party JavaScript libraries have been developed to enable this feature. This work employs Shaka Player~\cite{shaka_player}, a light and fully modular library that supports most operating systems and modern browsers.

\subsection{Basics of Named-Data Networking}
NDN is a large collaborative project to design and implement a scalable, resilient, and secure future Internet and is architected to distribute contents to millions of users efficiently.
NDN replaces IP addresses in network packets with content names and both requests (i.e., Interests) and responses (i.e., Data) carry the name of the solicited content rather than a source/destination address. Thus, instead of asking the network to deliver packets to a specific destination, content consumers request for contents and the network can retrieve them from anywhere. 
Names in NDN are hierarchically structured and composed of multiple components separated by ``/'', like \texttt{/ndn/video/playlist.m3u8}.
In NDN, every node maintains a table (called PIT) that keeps track of the Interests currently waiting for the corresponding Data packet. Returning Data packets follow the PIT entries, as if they were following a trail of breadcrumbs. 
This two-way Interest-Data exchange allows any NDN node to monitor the performance of its own data retrieval in terms of delay, jitter, loss, congestion, etc. and adapt itself to the network condition.
These fundamental changes enable a number of interesting features in the network like (i) native multicast and in-network caching, (ii) native multipath and multisource, and (iii) stateful and resilient packet forwarding. Moreover, all data packets in NDN are signed by their producers so that they are protected when stored and transferred. These features make NDN a natural fit for deploying a content distribution system, like video streaming service as the network can efficiently and reliably retrieve video segments from different sources through different paths while receivers can check the integrity of the Data packets, independently.

\subsection{Related Work}
A number of works in the literature designed and implemented native ICN applications~\cite{Jacobson'09, Gusev'15}. These works successfully re-engineered existing applications based on ICN paradigm so that the implemented/prototype applications can exploit the full architectural benefits of ICN. However, developing such applications can be very complicated and time-consuming. Moreover, native ICN applications are hard to deploy because they require end-users to install new software and applications. There is another line of works in the literature attempting to make existing applications work on top of ICN without modifying them~\cite {Linag2'18}. As the main advantages of these works, they require less development effort, and they can utilize most of ICN's benefits. However, end-users still cannot use these applications without manual configuration or installing new software. 

The current paper, on another hand, is an attempt to develop a service that end-users can use it with zero configuration. This work is less involving in terms of development effort. However, compared with a native application, it uses limited benefits of ICN. 

There are some other valuable works in the literature~\cite{Nguyen'16, Samain'17, Granl'13} that revealed several technical challenges and performance issues to adapt existing applications to the content-based paradigm, which are beyond the scope of this paper.

\section{NDN Video Streaming Service}

\begin{figure} 
    \centering
	\includegraphics[width=1\columnwidth]{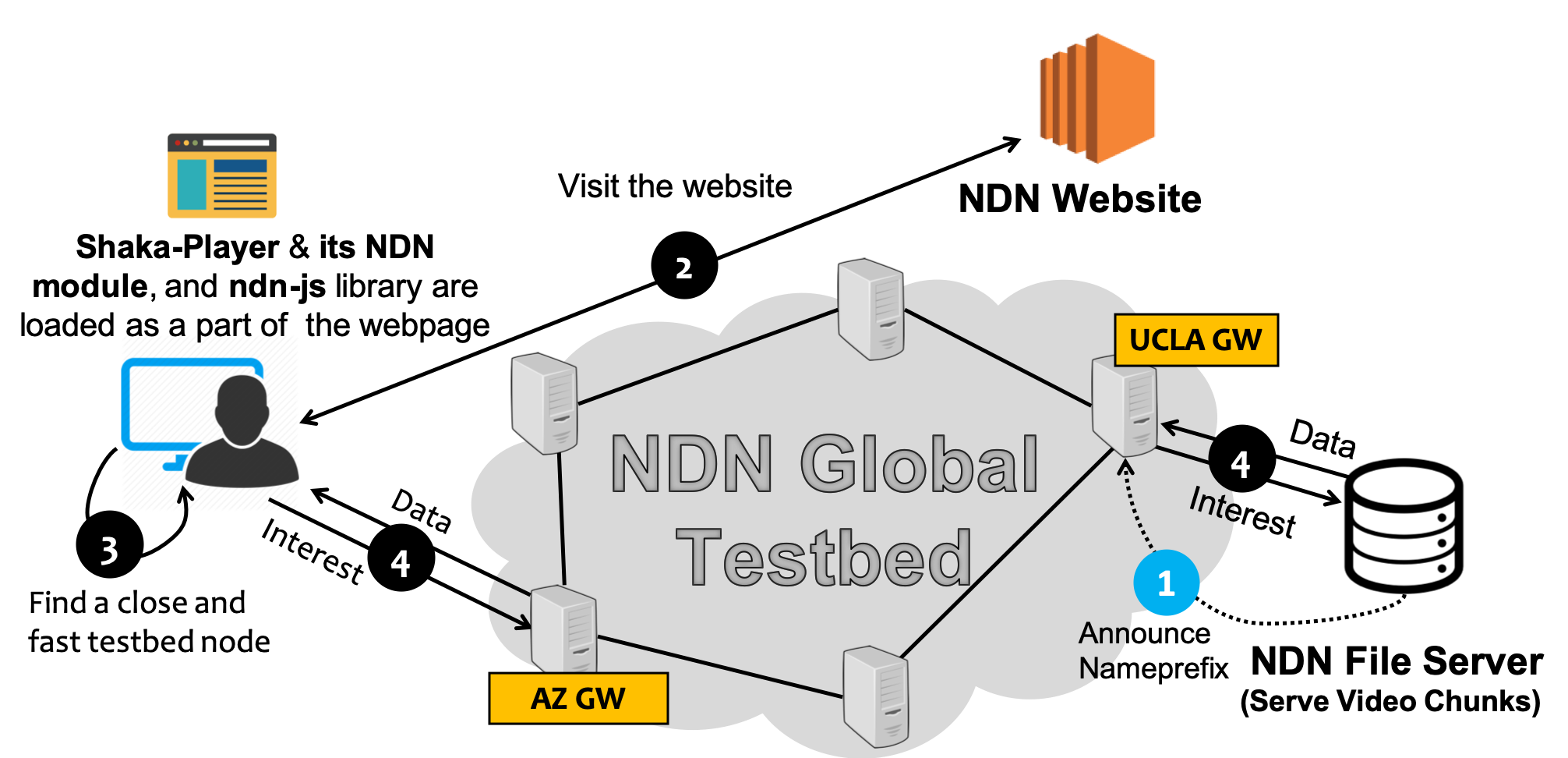}
	\caption{Overview of video streaming over NDN testbed}
	\label{fig:map}
\end{figure}


Fig.~\ref{fig:map} gives the system overview of our video streaming service and the specific steps involved. On the server side, video files are prepared and served as NDN packets; on the client side, browsers load a set of JavaScript libraries to support adaptive bit-rate streaming and NDN functionality; in the middle, the NDN testbed connects the server and client, and forwards NDN packets between them. 


\subsection{System Design}
\label{sub:architecture}


\subsubsection{Server Side}
To support various user devices and networks, for each video, we generate a set of different representations with different resolutions and encodings on the sever side.
We then package these sets of representations based on HLS standard, with a segment size of 2--4 seconds for a fair trade-off between encoding efficiency and the flexibility when adapting to bandwidth changes~\cite{segment_size}.



To stream a video in HLS format, all a browser needs is file transfer functionality over HTTP. To make the browser use NDN instead of HTTP, we need to divide each file into chunks on the server side, prepare these chunks as NDN data packets with proper signatures, and serve them when corresponding Interests are received.
We have developed a tool to partition each representation of a video into a series of chunks, sign them, and store them in Mongo database. 
Due to the space limitation, we do not detail the design of namespace for Mongo database and refer the readers to \cite{ivisa-website}.

\begin{figure} 
	\includegraphics[width=0.95\columnwidth]{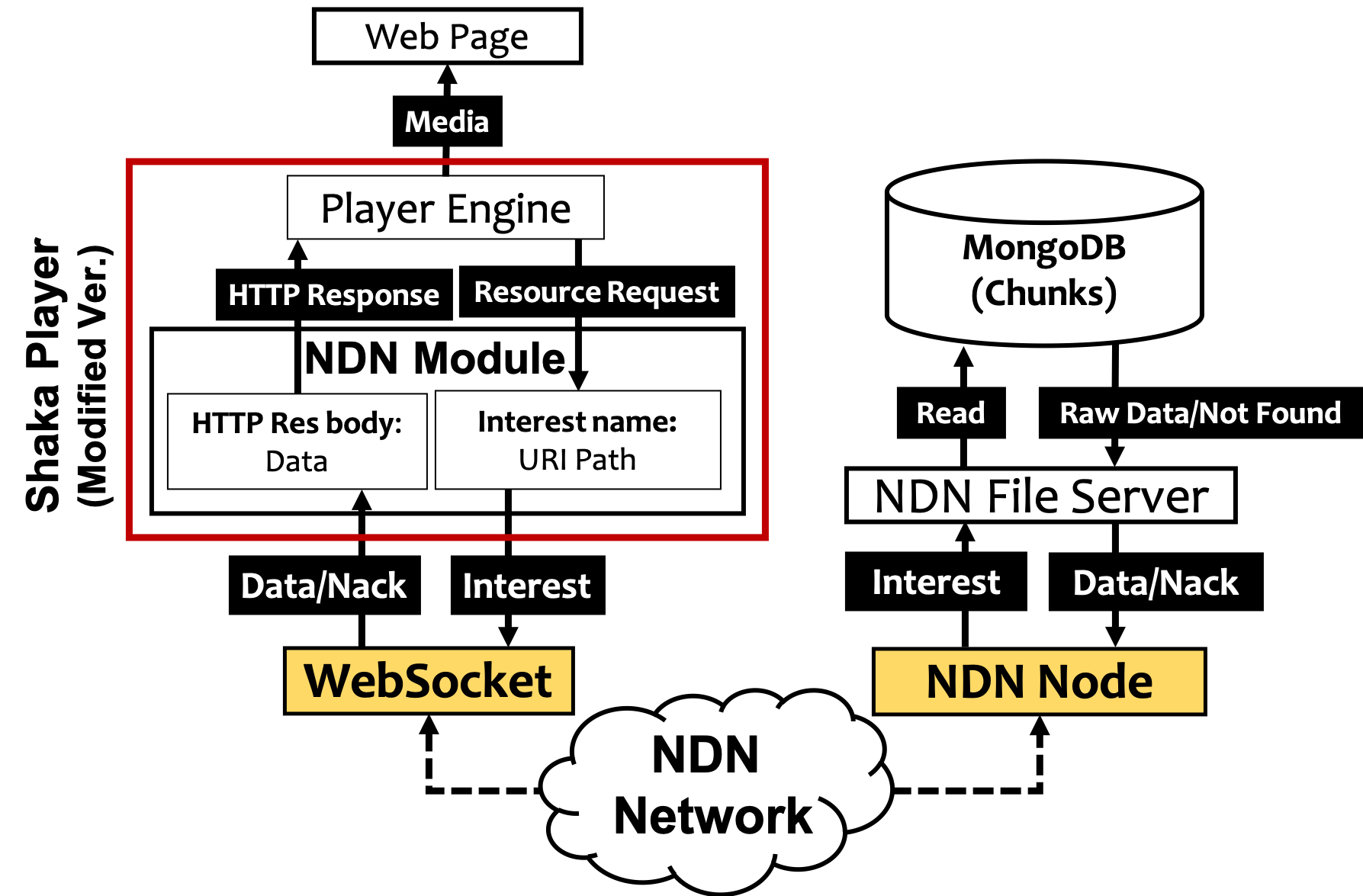}
	\caption{Architecture of client and server sides of our NDN video streaming service}
	\label{fig:videostreaming}
\end{figure}

Each chunk can then be served in its entirety within a Data packet by issuing an Interest for the said Data packet. 
The naming of these data chunks follows a convention like \texttt{ /ndn/web/video/<video-file-name>/<version> /<chunk-number>}. 
As shown on the right side of Fig.~\ref{fig:videostreaming}, we have developed an NDN file server to serve the stored chunks. Upon receiving an Interest, the file server reads the corresponding chunk from the database, packages it as a Data packet, and sends it out. If the requested chunk does not exist, a NACK will be sent back.

\subsubsection{Client Side}
\begin{figure} 
	\includegraphics[width=0.93\columnwidth]{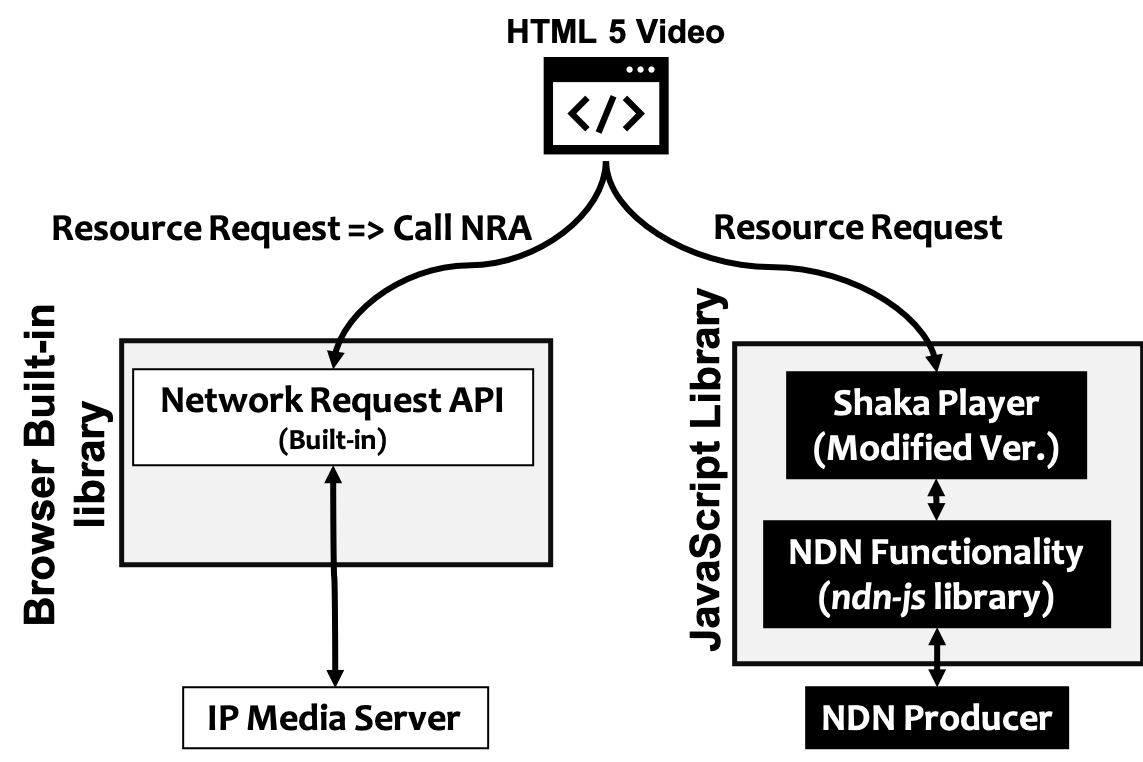}
	\caption{How browsers stream videos over IP vs NDN}
	\label{fig:browser}
\end{figure}

Employing JavaScript technology for video streaming inside web browsers provides a solution with zero client-side configuration, cross-platform support, and transparent service. Fig.~\ref{fig:browser} compares how existing technology streams a video over HTTP versus how NDN does the job. Here, each video segment is a resource for which the video player makes a network request call. A resource request is represented by a URL, and the resource can be any file (e.g., playlist, video, or audio segment) for playing back the solicited video. To resolve a resource request, browsers provide applications with Network Request APIs (NRA) (see the left box in Fig.~\ref{fig:browser}). NRA sends an HTTP request over TCP to the destination to fetch the requested resource. Upon receiving the resource, it forms a proper HTTP response containing the received data and sends it back to the caller (e.g., video player). 

To make the video player (or any HTTP-based application in general) use NDN for network communications, we need to override the handler of resource requests.
We can directly override the built-in NRA of browsers by loading custom JavaScript code on the client side to get NDN functionalities. This leaves the application intact while redirecting its network request calls to a custom JavaScript code. From there on, NDN takes care of fetching the solicited resource and sends a proper HTTP response back to the video player upon retrieving the resource. However, this requires our code to be able to translate between HTTP packets and NDN packets, which is a significant challenge due to the large number of features and header options of HTTP. 
Thus, as the right side of Fig.~\ref{fig:browser} shows, instead of employing browsers' built-in NRA, our video application makes all its network requests to a third-party JavaScript library, i.e., Shaka Player. To let this library employ NDN technology to handle network requests, we have developed a new network module that employs NDN-JS and added it to Shaka Player by which all network requests will be resolved over NDN protocol --- that is why we refer to it as a \textit{modified version} in Fig.~\ref{fig:browser}.

As shown on the left side of Fig.~\ref{fig:videostreaming}, when the NDN network module loads, it connects to an NDN node (probably one of the NDN testbed nodes) via WebSocket.
After receiving a resource request from video player, the NDN module starts expressing Interests to fetch the chunks of the solicited resource.
A resource request contains \textit{protocol} and \textit{URI path}. For both HTTP and HTTPS protocols, the request is redirected to the NDN module as we use NDN protocol for client-server communications. Usually, the client will retrieve the playlist file first, then fetches individual video and audio chunks. A basic name discovery mechanism is involved in retrieving each file. For example, to fetch the playlist file, the client sends the first Interest with name \texttt{/ndn/web/video/foo/playlist.m3u8}, but it does not know the version number of the file. Upon receiving this Interest, the file server returns a Data packet containing full name of the solicited content (including version and chunk number) as well as the content of the first chunk of \texttt{playlist.m3u8} file. Then, the client learns the version number and expresses complete names to retrieve the remaining chunks.

\subsubsection{Network}
The NDN testbed is an overlay network comprised of over thirty-two member institutions, mostly universities, from all around the world in support of building a world-wide research network for development and testing of emerging ICN protocols, services, and applications. By connecting a new file server to the testbed, the nameprefix of file server's contents will be propagated to the entire testbed. When Interests for this nameprefix arrive, the testbed nodes know how to forward them. If there are multiple sources (producer or temporary caches) at different locations, the testbed routing protocol and forwarding strategy will decide the best way to forward Interests. For routing security purpose, each nameprefix announcement needs to be properly authorized. Therefore, the content providers need to obtain a valid certificate for the nameprefix (e.g., \texttt{/ndn/web}) from the testbed administrator before making the announcement.

We employ an HTTP-based service called ``Find Closest Hub'' (FCH), running on the NDN testbed, that allows any client to find geographically nearby testbed nodes. However, the closest node does not necessarily provide the best performance due to different reasons, e.g., the node's workload or its network condition. Thus, in our service, we test resolved nodes from the FCH service and figure out how fast each one responds on client side, and finally choose the one with the shortest response time. (In the rest of the paper, we refer to this node as gateway). The client then makes a WebSocket connection to the chosen testbed node and uses it for all NDN communication. 
Putting everything together, Fig.~\ref{fig:map} illustrates all the steps involved in this video streaming service. First, video files are prepared at the server and their name prefixes are announced into the testbed. Second, the client uses a web browser to access a web page that contains all required JavaScript libraries.
Third, the NDN network module calls FCH and provides the client with a few number of nearby testbed nodes. Fourth, the client connects to its gateway and upon starting the video, the video player sends resource requests, which are turned into NDN Interests, and when NDN Data come back, they are returned to the player as video files. Note that the entire process is transparent to the end-user.

\begin{figure}
\centering
    \includegraphics[width=8.5cm]{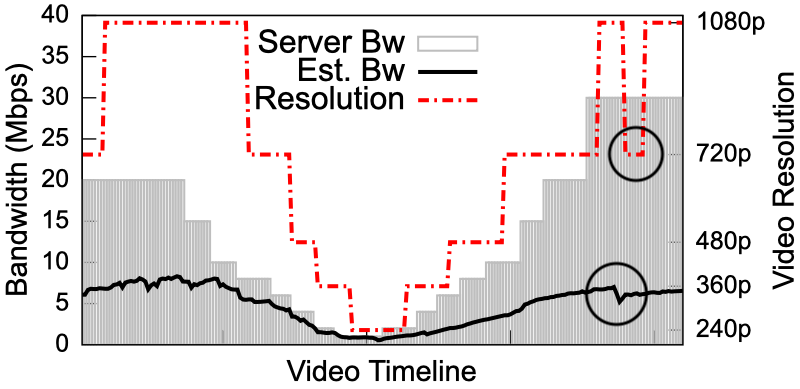}
    \caption{Adaptive bit-rate feature of the service while throttling the egress bandwidth of the file server}
	\label{fig:abr}
\end{figure}

\section{Evaluation}

In this section, we first validate the adaptive bit-rate feature of the service. We then present the evaluation results on client, network, and server sides, separately, providing a clear understanding of the performance of each side: (1) on the client side, we measure QoE in terms of experienced video quality, number of re-bufferings during video playback, and video startup delay; (2) on the network side, we focus on Interest-Data RTT and jitter while revealing the role of in-network caching; and, (3) on the server side, we show the server's contribution to overall content retrieval delay by measuring how fast it responds to incoming Interests. 
It is worth mentioning that all performance measurements in this section are reported by our statistics collector tool which is not presented in this paper due to the page limit. We refer the reader to \cite{ivisa-website} for design and implementation details of this tool.

We evaluate the performance of the service in two different scenarios: (1) \textit{no-cache:} when no part of the video has been cached in the network, and (2) \textit{with-cache}: when some parts of the video have been already cached in the network.

\begin{figure*}
\begin{minipage}[]{0.6\linewidth} 
\subfloat[]
	{                \hspace{-0.8cm}\includegraphics[width=5.6cm,height=3.2cm]{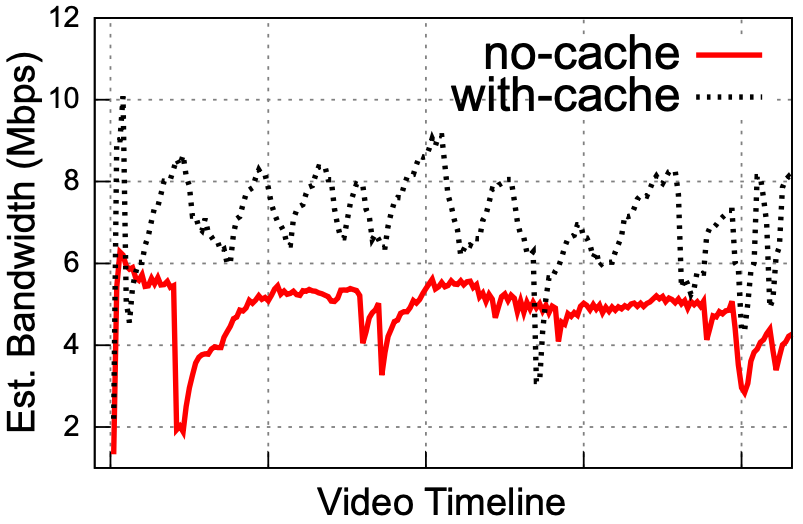}}
	\hfill
	\subfloat[]
	{                \includegraphics[width=5.6cm,height=3.2cm]{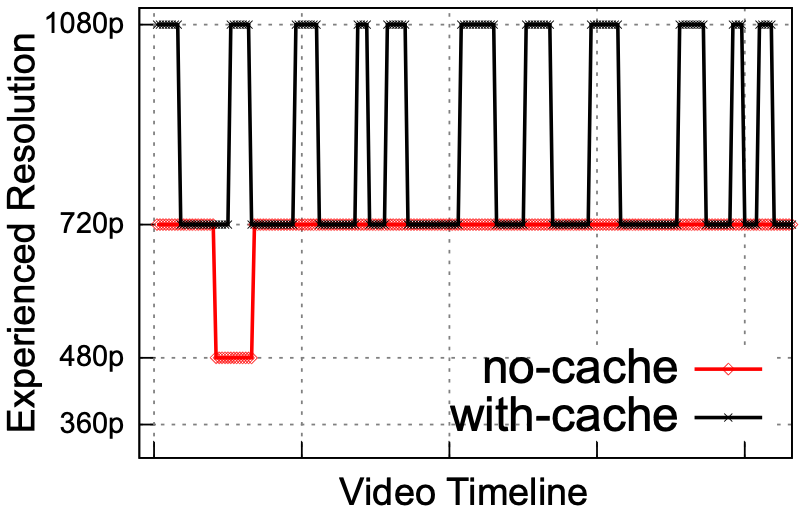}
	} \caption{(a) Experienced bandwidth and (b) video quality in client side during \\video playback}
	\label{fig:QoE}        
\end{minipage}
\centering
\begin{minipage}[]{0.3\linewidth} 
    \includegraphics[width=6.8cm]{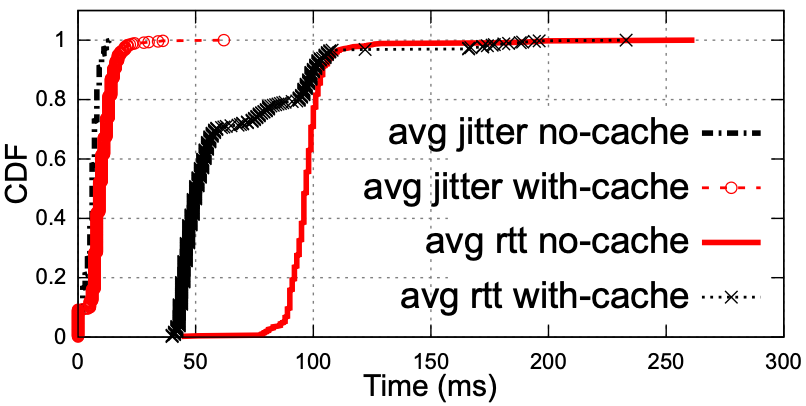}
    \caption{Average Interest-Data RTT and jitter for each file during video playback}
	\label{fig:rtt}
\end{minipage}
\end{figure*}

\subsection{Experimental Setup}
The core network in our experiments is the global NDN testbed, and our file server that serves NDN website videos resides in Portland. The end-user is in Arizona and watches a number of videos on the NDN website. Due to the space limit, we only report the average results for single-video scenarios for one week-long period.


For each video on NDN website, our server provides the end-users with five different representations to support different connection qualities. Here are the resolution and minimum required bandwidth of each representation: 240p (very poor): \texttt{0.6Mbps} || 360p (poor): \texttt{0.9Mbps} || 480p (medium): \texttt{1.8Mbps} || 720p (HD): \texttt{3.3Mbps} || 1080p (full HD): \texttt{6.3Mbps}.



\subsection{Results}

\subsubsection{Adaptive Bit-Rate Streaming}
An essential feature of our service is the support of adaptive 
streaming, enabled by employing a third-party JavaScript library. To validate this feature, Fig.~{\ref{fig:abr}} shows the consumer's behavior while the egress bandwidth of the file server is throttled. To disable the contribution of in-network caching in this scenario, the end-user establishes a direct tunnel to the server. The player keeps looking at content retrieval performance (e.g., RTT, number of timeouts, etc.) and switches between available bit-rates (i.e., video resolutions) based on network bandwidth estimation at a given moment in time. 

We set the server's egress bandwidth to 20Mbps and decrease it step by step. After startup, the player switches to full HD resolution and stays with it until the server bandwidth reaches 8Mbps. From there on, the video resolution downgrades until it switches to 240p, where the server's and player's estimated bandwidths drop down to 1Mbps and 0.8Mbps, respectively. On the other hand, by increasing the server's egress bandwidth, the player improves the resolution. At one point (shown by circles), the video resolution drops to 720p, which is because of network congestion effect on player's bandwidth estimation.

\subsubsection{Client-Side Evaluation}
Three performance metrics play vital roles in measuring the QoE on the client side: (1) \textit{startup delay,} the amount of time that the end-user waits until the video starts playing, (2) \textit{number of re-buffering}, the number of stops/stalls during video playback, and (3) \textit{video quality}, the video resolution that the end-user experiences. 

The video startup delay in \textit{no-cache} scenario is 4.19 seconds. The startup delay includes the delay of loading necessary resources (including JavaScript libraries) and buffering enough amount of video (i.e., 2--4 seconds). According to~\cite{Krishan13}, the end-users prefer a startup below 2 seconds, and each incremental delay of 1 second can result in a 5.8\% increase in the user abandonment rate. Thus, there is still room to improve the service and NDN system. Moreover, in none of the runs, the end-user has experienced re-buffering during video playback. 



\textit{In-network caching role}---We also investigate how in-network caching can affect the user's QoE. Among all the runs, we have chosen the one that fairly represents the average behavior of the system in terms of video quality experienced by the end-user and Fig.~{\ref{fig:QoE}} compares \textit{no-cache} and \textit{with-cache} scenarios for that run. For all \textit{with-cache} runs, we made sure that the network has already cached 720p version of videos before the end-user in Arizona starts watching the videos.
Thus, in \textit{with-cache} scenario, when the end-user starts to watch the videos, the player fetches 720p version from the gateway (Arizona hub). Because the channel between the end-user and its gateway is pretty fast, the player estimates that the network is good enough to switch to a better video resolution, asking for a full-HD version. However, the full-HD version is not cached in the network. Therefore, the player senses the channel to the producer (file server). This time, it experiences a longer delay and lower network throughput. Thus, the bandwidth estimation decreases and the player switches to 720p video. Looking at the figures, we can see that the end-user's estimated bandwidth and experienced video quality oscillate, which is a direct result of in-network caching. 

Although comparing \textit{with-cache} and \textit{no-cache} scenarios in Fig.~{\ref{fig:QoE}} shows that caching improves the video quality experienced by the end-user, in-network caching misleads player's bandwidth estimation as it causes the player to sense more than one channel during video playback. This problem is discussed in detail by~\cite{Granl'13, Nguyen'16}. Besides, according to our results, the cache hit ratio in \textit{with-cache} scenarios is less than 53\% (while 92\% of 720p version of the video had been cached in the network). This reveals that the end-user does not fully exploit the in-network caching potentials. Although it is beyond the scope of this paper and part of our future work, we believe that employing a layered video encoding technique can address this problem as by using this technique, the end-user retrieves 720p video from the gateway and requests only the remaining enhancement layers for 1080p representation from the server. In this way, not only the end-user can experience full-HD video for probably the entire video playback, but also the cache hit ratio increases (and server workload decreases) by 92\% instead of 53\%. 

It is worth noting that utilizing in-network caching improves the video startup delay by 1 second and the end-user experienced no re-buffering during video playback.

\subsubsection{Network-Side Evaluation}
In this section, we study the network-related performance metrics. As mentioned earlier, each video is composed of several files (e.g., video and audio segments, playlist files, etc.), and each file may include one or more chunks. For a given file, we calculate the average retrieval delay of its chunks and report it as average Interest-Data RTT (or RTT for short). For a particular chunk, RTT counts the time between the latest sent Interest to the earliest received Data packet for the said Interest. Fig.~{\ref{fig:rtt}} depicts cumulative distribution function (CDF) of the average RTT of all retrieved files. In \textit{no-cache} scenario, the average RTT of 90\% of files falls in the range of 78ms to 104ms. (According to~\cite{Cisco_QoS}, for smooth video playback, RTT should be kept below 150ms, which shows that our service's latency is acceptable.) On the other hand, in \textit{with-cache} scenario, the average RTT of 90\% of files falls in the range 40ms to 100ms. This improvement is a direct result of retrieving over 50\% of the video from the end-user's gateway.

Another essential metric to consider is jitter. It shows delay variance of successive packets and is important as any significant changes in jitter can cause re-buffering, and thus, a poor performance. Fig.~{\ref{fig:rtt}} shows that the average jitter of all files in \textit{no-cache} scenario is less than 13ms, which meets the basic standards of video-on-demand service. (According to~\cite{Cisco_QoS}, keeping jitter below 30ms leads to a fluent video playback). On the other hand, in \textit{with-cache} scenario, the percentage of reported files with jitter 13ms is 81\%. Moreover, based on our results, in-network caching can cause the average jitter to exceed 30ms at some points. This is because of fetching the content from two different nodes (i.e., the gateway and server) during watching the video that magnifies the delay changes. Although the effect of jitter on video playback is mitigated by buffering, we see that in-network caching can potentially worsen the jitter of a video streaming service. 
A viable solution to this issue is to pre-fetch Data packets of the solicited contents at the gateways. More specifically, upon receiving the very first Interests for a content, the end-user's gateway starts fetching a sufficient number of upcoming Data packets to satisfy the future end-user's Interests from its cache, right away. This approach minimizes jitter by bounding the Interest-Data delay to the RTT between the end-user and its gateway. NDN fortunately comes with a module, called \textit{forwarding strategy}, which allows researchers and developers to rule whether, when, and how to forward the Interests on each network node. We believe an interesting way to realize this solution is to design a new forwarding strategy equipped with a pre-fetching mechanism.

\subsubsection{Server-Side Evaluation}
On the server side, an important metric to consider is the server's response time that shows the server's contribution to content retrieval delay. It includes the time for processing a received Interest, reading the associated bytes from disk, creating a Data packet, and sending it out/back. 
Our results show that over 98\% of received Interests during video playback for all runs are satisfied in less than 5ms. This shows a reasonable implementation of our file server and clarifies that the server is not the bottleneck in the content retrieval pipeline.

\section{Discussion}

\textbf{\textit{Running existing applications over NDN}}: As a critical deployment challenge, ICN/NDN needs the existing applications to be rewritten or modified in order to understand the NDN protocol and fully receive its architectural benefits. From-scratch and proxy-based solutions are two traditional lines of works that try to address this issue, both of which have an involving development process and require end-users to install new software or apply manual configurations. This paper, on the other hand, opens a new line of work in the literature and introduces a new method of thinking to run existing applications/services over NDN (1) without modifying the application; we propose to design and implement an NDN module to handle all application's network interactions on the front-end, (2) without involving the end-user; we propose to employ zero-configuration technologies like JavaScript on the front-end to enable NDN functionalities on-the-fly, and (3) with exploiting NDN's architectural benefits; we propose to design and implement the application's back-end over NDN.\vspace{0.2cm}

\textbf{\textit{NDN's feasibility in the real world}}: Since the very beginning, the NDN community has faced a major question yet with no answer: \textit{``Can NDN work in the wild?''} This work, for the first time to our best knowledge, gives a clear answer YES to this question. While NDN has been mostly an academic research project for the future Internet, the current work brings it to everyday users and showcases NDN's capability to work with real-world settings. It is worth mentioning that this system has been serving over hundreds of gigabytes of the official NDN website to thousands of daily Internet users during the past few months.\vspace{0.2cm}

\textbf{\textit{Open problems}}: By involving a large number of users, this work will significantly help the NDN community to find NDN's design and implementation problems, opening up an avenue for future research. This paper reveals two interesting problems caused by in-network caching for time-sensitive applications: (1) traditional ABR mechanisms cannot be directly used in NDN as they have been designed for single-channel networks while in-network caching makes NDN a complete multi-channel network. We believe employing a layered video encoding technique as well as designing a multi-channel ABR algorithm are two approaches to address this issue, and (2) in-network caching intrinsically causes high delay variances (jitter) on the end-user side. We suggest designing a new forwarding strategy with a pre-fetching mechanism to alleviate this problem.

\section{Conclusion}
This work presented the system design of an adaptive video streaming service over NDN as the first public ICN/NDN service for daily Internet users and detailed its deployment on the global NDN testbed. Our experimental results validated the proper design and implementation of our software and a reasonable QoE of the service over the global NDN testbed, using the official NDN project's website video contents.

\vspace{1cm}
\bibliographystyle{IEEEtran}
\bibliography{references} 

\end{document}